\documentclass[sigconf]{acmart}
\usepackage{color}
\usepackage{flushend}
\usepackage{graphicx}
\usepackage{booktabs}
\usepackage[htt]{hyphenat}
\usepackage[]{xcolor}
\definecolor{midnightgreen}{rgb}{0.0, 0.29, 0.33}

\newcommand{\jeff}[1]{} %{\jeffgrumbler{Jeff}{#1}} %

\begin{document}
\title{CAsT 2019:  The Conversational Assistance Track Overview}
%
%\titlerunning{Abbreviated paper title}
% If the paper title is too long for the running head, you can set
% an abbreviated paper title here
%
% \author{First Author\inst{1}\orcidID{0000-1111-2222-3333} \and
% Second Author\inst{2,3}\orcidID{1111-2222-3333-4444} \and
% Third Author\inst{3}\orcidID{2222--3333-4444-5555}}
% %
% \authorrunning{F. Author et al.}
% First names are abbreviated in the running head.
% If there are more than two authors, 'et al.' is used.
%
% \institute{Princeton University, Princeton NJ 08544, USA \and
% Springer Heidelberg, Tiergartenstr. 17, 69121 Heidelberg, Germany
% \email{lncs@springer.com}\\
% \url{http://www.springer.com/gp/computer-science/lncs} \and
% ABC Institute, Rupert-Karls-University Heidelberg, Heidelberg, Germany\\
% \email{\{abc,lncs\}@uni-heidelberg.de}}
%

\author{Jeffrey Dalton$^1$, Chenyan Xiong$^2$, and Jamie Callan$^3$}
\affiliation{University of Glasgow$^1$,  Microsoft Research$^2$, Carnegie Mellon University$^3$ \\
\texttt{jeff.dalton}@glasgow.ac.uk, \texttt{chenyan.xiong@microsoft.com}, \texttt{callan}@cs.cmu.edu
} 
\maketitle              % typeset the header of the contribution
\begin{abstract}
We provide an overview of the task, dataset, and results of the first year of TREC Conversational Assistance Track (CAsT).  We define the conversational search tasks and how the provided datasets are constructed. Twenty one teams participated submitting runs exploring the use of natural language understanding, conversational retrieval models, as well as neural passage ranking methods. We see that ...
\end{abstract}

% \textcolor{green}{Jamie: These text margins are ridiculously wide.  I prefer 1 inch.}
% \cx{Are there any format requirements for the overview paper?}

\section{Introduction}
% \cx{Will need to update this to ``overview'' from ``proposal''}

The importance of conversation and conversational models for complex information seeking tasks is well-established within information retrieval, initially to understand user behavior during interactive search \cite{belkin95,solomon97} and later to improve search accuracy during search sessions \cite{carterette15}. The rapid adoption of a new generation of \textit{conversational assistants} such as Alexa, Siri, Cortana, Bixby, and Google Assistant increase the scope and importance of conversational approaches to information seeking and also introduce a broad range of new research problems \cite{swirl2018}.

%The recent investigation, nonetheless, was mainly in the task-oriented or chit-chat space when studying conversational assistants, while how the conversational assistants and the search engines can better support information seeking conversations, is much under explored.

The TREC Conversational Assistance Track (CAsT) is a new initiative to facilitate Conversational Information Seeking (CIS) research and to create a large-scale reusable test collection for conversational search systems.
We define it as a task in which effective response selection requires understanding a question’s context (the dialogue history). It focuses attention on user modeling, analysis of prior retrieval results, transformation of questions into effective queries, and other topics that have been difficult to study with previous datasets.

%The task is, given the previous turns in the conversational context, to retrieve passage-length texts that directly satisfy a person's information need and/or serve as input to more complex downstream steps.

To make this tractable and reusable for the first year of CAsT, we begin with pre-determined conversation trajectories and passage responses. 
Our target conversations include several rounds of utterances that are coherent in topic and explore relevant information.
The primary initial focus is on system understanding of information needs in a conversational format and finding relevant passages leveraging conversational context.

The long-term vision of CAsT is to allow natural conversions with 
mixed-initiative, where the system performs a variety of information actions \cite{radlinski2017theoretical}, e.g., providing information (INFORM), asking clarifying questions (CLARIFY), leading conversations with more interactions (SUGGEST), and others. For the first year we focus on context understanding and use simple INFORM actions, where systems return text passages to the user.
In the future, we plan to explore richer sets of information actions, richer response formats, and more interactions between users and conversational agents.

\section{Task Description}
CAsT defines conversational search as an information retrieval task in the conversational context.
The goal of the task is to satisfy a user's information need, which is \emph{expressed or formalized through turns in a conversation}. 
The response from the retrieval system is not a list of relevant documents. Instead the response is limited to \emph{brief text passages (approximately 1-3 sentences in length)} suitable for presentation in a voice-interface or a mobile screen.
%The key difference is that the information need is more sophisticated or is not initially fully specified by the user, thus requires a sequences of conversational queries $\{q_1,...q_t,...,q_T\}$ to represent.

\textbf{Task Definition.}
The task in Year 1 focuses on candidate response retrieval for information seeking conversations. 
Our goal is to create a low barrier to entry as well as keep the task simple for the purpose of creating a reusable collection. 
Specifically, given a series of natural conversational turns for a topic, $T$, with utterances ($u$) for each turn $T=\{u_1,...u_i...u_n\}$, the task is to identify relevant passages $P_i$ for each turn (user utterance) $u_i$ to satisfy the information needs in round $i$ with the context in round $u_{<i}=u_1:u_{i-1}$.

\begin{table}[t]
    \centering
     \caption{CAsT Training Topic 18.    \label{tab:topic_train}}
\begin{tabular}{ll}
% \multicolumn{2}{l}{Number: 18} \\ 
\hline
\multicolumn{2}{l}{\textbf{Title}: Uranus and Neptune} \\
\multicolumn{2}{l}{\textbf{Description}: Information about Uranus and Neptune.} \\ \hline
\textbf{Turn} & \textbf{Conversation Utterances} \\ \hline
1 &	Describe Uranus. \\
2 &	What makes it so unusual? \\
3 &	Tell me about its orbit. \\
4 &	Why is it tilted? \\
5 &	How is its rotation different from other planets? \\
6 &	What is peculiar about its seasons? \\
7 &	Are there any other planets similar to it? \\
8 &	Describe the characteristics of Neptune. \\
9 &	Why is it important to our solar system? \\
10 &	How are these two planets similar to each other? \\
11 &	Can life exist on either of them? \\
\hline
\end{tabular}
\end{table}

To construct this task, we start with a selection of open-domain exploratory information needs and create  conversational topics $T$. Then we use the passage collection $C$ to provide candidate response passages for those topics.

\textbf{Information Needs.}
We semi-manually constructed exploratory information needs (topics) from the combination of previous TREC topics (Common Core, Session Track, etc.), MS MARCO Conversational Sessions (described in Sec~\ref{sec:pretrain}), and our own interests and experience. The information needs were selected to ensure complexity (requiring multiple rounds of elaboration), diversity (across different information categories), open-domain (not requiring expert domain knowledge to access), and answerable (sufficient coverage in the passage collections). 

\textbf{Conversational Sequences}.
We manually created the sequences of conversation utterances for each turn in a topic. In general, we started with a general introduction of the topic and then manually formulated exploratory information seeking trajectories. For the re-usability of the topics for year one we ensured that later turns only depended on the previous utterances, not on system responses (an area for future work). 
%The trajectories are curated based on our understanding of the task and a common web user's behavior. They are also inspired by the original search sessions of the topic, from Session Track or MS MARCO, if exists.

When curating the conversational trajectories multiple sources of information are used. The MS MARCO search session data is one input. Query suggestions from commercial search engines (Google and Bing) and specifically the natural language questions from the ``People Also Ask'' feature in Google and Bing are used.  These questions are similar to the questions released in the Google Natural Language Questions dataset \cite{google-natural}. 

The conversational sequences are written to mimic ``real'' dialogues. Namely, we ensure that there are coherent transitions between turns. We also introduce common conversation phenomena including coreference and omission. Comparisons between various subtopics are also introduced where relevant. To focus on long-form dialogue most topic turns require more than a short answer response (i.e., a simple factoid response is insufficient).

An example topic from the released training set is shown in Table~\ref{tab:topic_train}.
For the first year of the track, we developed 30 training and 50 evaluation topics, each with about ten turns. The evaluation conversations cover a diverse range of open-domain topics.

\textbf{Passage Collection.}
The corpora used are passages from MS MARCO\footnote{\url{http://www.msmarco.org/}}, the TREC Complex Answer Retrieval Paragraph Collection \cite{dietz2018trec}, and the Washington Post\footnote{\url{https://trec.nist.gov/data/wapost/}} collections. 

% For a collection, we will use a fixed collection of large open-domain text passages. We plan to use passages from Wikipedia, more specifically, paragraphs from Wikipedia similar to those used in the TREC CAR Track.
The TREC CAR (Wikipedia) paragraphCorpus V2.0\footnote{\url{http://trec-car.cs.unh.edu/datareleases/}} is used, which consists of all paragraphs from Wikipedia `16. Note that this corpus has been deduplicated. It contains approximately 30 million unique paragraphs. Refer to the TREC CAR Overview~\cite{dietz2018trec} for details on how this corpus was created.  

% This is a proven collection of data that includes diverse information topics as well as entity representations. 

MS MARCO also has ~1M real search queries each with 10 passages from top ranked results, resulting in a pool of approximately ~8 million passages. 
A passage's metadata includes the source URL, the MARCO queries associated with it, and relevance labels for adhoc passage retrieval. The MARCO collection does contain near duplicates. 

%These Wikipedia passages include various aspects of entities. They are a reasonable starting point, and their feasibility in satisfying complex information needs has been demonstrated in the CAR Track.

The Washington Post Collection (WaPo) collection was also initially used and was included in the submitted runs. When de-duplicating the WaPo corpus an error in the process led to ambiguous document ids. As a result, the final assessments are restricted to MS MARCO and CAR passages. The MS MARCO and CAR passages combined provided a sizable and proven collection of passages. The WaPo documents are filtered from all the runs for pool creation and evaluation (this removed less than 5\% of results returned by systems). 

%  The organizers provided a list of duplicate and near-duplicate document ids to be filtered from MS MARCO and WAPO, to reduce assessing costs and duplication among retrieved results.  However, an error in the near-duplicate processing of the WAPO corpus rendered the WAPO document ids ambiguous, thus the final judgments were restricted to just MS MARCO and WAPO passages.

\section{Resources}
\label{sec:pretrain}
Beyond the conversational topics and passage collections, the organizers also provided  additional resources to track participants and release those feasible ones to the public for future CIS research.

\jeff{We could restructure this a bit.  Training data, NLP labels, software tools, auxiliary data (dedup), etc...}

\subsection{Training Data and Manual Annotations.}
We curated and provided three resources for model training: training topics with incomplete manual judgments, MS MARCO conversational search sessions, and manually rewritten topics. We also provided near-duplicate files of the passage collections.  

\textbf{Training data.}
The organizers created thirty training topics. Five of these topics have manually created relevance assessments developed by a CMU PhD student (approximately 50 turns). Relevance assessment for these was performed on a different (compressed) three-point relevance scale. 

\textbf{External Data.}
% Neural models often require large scale training data.
% and our training
Building on MARCO and TREC CAR collections allowed this track to share previous existing relevance labels for the non-conversational topics. These labels could be used by participants to train single-shot relevance. 
%The passage ranking task of MS MARCO serves as our recommended weak supervision dataset for CAsT systems. It includes millions of training labels which are sufficient to train single-round deep neural ranking models. 

For CAsT the organizers also created an extension of the MS MARCO collection, the MS MARCO Conversational Search Session dataset\footnote{https://github.com/microsoft/MSMARCO-Conversational-Search}. The Conversation Search Sessions are constructed by aligning the one million released MS MARCO queries to Bing search sessions, to simulate actual sessions in search logs.
The alignment was conducted using the Generic Intent Encoder that maps queries with similar search intents together~\cite{genencoder}.
We first obtained all the encodings of MS MARCO queries and built an approximate nearest neighbor index (ANN)\footnote{https://github.com/spotify/annoy}. Then for each Bing search session, we ran its queries on the ANN index and replaced the Bing query with the most similar MS MARCO query if their cosine similar was greater than 0.85 (considered to be paraphrases). If such a MS MARCO query did not exist, the query was discarded from the session. We kept those sessions longer than three queries that are also topically coherent following gen encoder's definition~\cite{genencoder}. 
The result is a publicly available dataset of realistic information seeking sessions.

\textbf{Manual Conversation Rewrites.}
As shown in Table~\ref{tab:topic_train}, the topic utterances include natural phenomena including coreference and omission. To facilitate assessment and research, we manually created an annotated dataset for the training and evaluation topics to resolve ambiguity and implicit conversational context.
The manually rewritten utterances (``resolved'') contain all of the information required to represent the single turn of the underlying information need. 

Each utterance was rewritten by two organizers.  The results were compared and the disagreements adjudicated to a canonical form. 
%There was not much disagreement because the organizers agreed previously on the information needs when curating the conversations. 
On average it took approximately 5-10 minutes to rewrite a topic (ten turns on average), indicating this is non-trivial even for those familiar with the topics.

\textbf{Passage Collection Deduplication.}
\jeff{A bit more detail on the deduplication process would be helpful here, parts are a bit vague.}
Early results found that both the MARCO and WaPo corpora contain a significant number of near-duplicate paragraphs. The organizers ran near-duplicate detection to cluster results; only one result per duplicate cluster was evaluated. The organizers recommended to participants that they remove duplicates (keeping the canonical document) from their indices.

The MARCO near-duplication algorithm grouped passages based on their URLs.  Within each URL group, passages were first sorted by their length in descending order.  Pairwise matches between the passages in the group were calculated. A pairwise match was defined as the total percentage of matching words in the smaller passage with respect to the longer passage in the input pair. If this percentage match was greater than 95\%, then the ID of the smaller passage was added to the near-duplicate dictionary. The IDs in the final duplicate dictionary were then mapped back to the MS-MARCO ranking corpus IDs (based on a prior alignment of IDs between the two corpora).

A similar procedure was used for WaPo, but the organizers discovered that the near-duplicate algorithm was run on an outdated version of the WaPo collection and could not be corrected easily. 

\subsection{Software Tools}
The track provided various software tools to support the development of CIS systems. 

\textbf{CAsT Topic Tools.\footnote{https://github.com/grill-lab/trec-cast-tools}}
These tools are publicly available. The tools include sample code to load the conversation topics in both Python and Java. The topic files are available in multiple formats including JSON, text, and Google Protocol Buffers.  The protocol buffer format is the canonical representation.

\textbf{Indri Baseline Retrieval System.}
The organizers provided web access to an Indri search engine for the CAsT corpus.\footnote{\url{http://boston.lti.cs.cmu.edu/Services/treccast19/} and \url{http://boston.lti.cs.cmu.edu/Services/treccast19_batch/}}  During indexing, Krovetz stemming \cite{krovetz1993viewing} was applied and stopwords were retained.  CAR passages were indexed with title and body fields. \jeff{CAR paragraphs don't have titles, so what is this?}  MARCO and WAPO passages were indexed with only body fields.  Near-duplicates were discarded from the MARCO and WAPO collections, as described above.  Interactive and batch search were supported, with up to 1,000 results returned.

\section{Evaluation Methodologies} 

\begin{table}[t]
    \centering
    \caption{Judgment statistics}\label{tab:judgmentsstat}
    \begin{tabular}{l|c}
    \hline \hline
    Topics & 20 \\
    Turns & 173 \\
    Assessments   & 29,571  \\
   % Total relevant (binary) & 8,120\\
    \hline
    Fails to meet (0) & 21,451\\
    Slightly meets (1) & 2,889\\
    Moderately meets (2) & 2,157\\
    Highly meets (3) & 1,456\\
    Fully meets (4) & 1,618\\
    \hline \hline
    \end{tabular}
\end{table}

%The twenty evaluation topics are: 31 32 33 34 37 40 49 50 54 56 58 59 61 67 68 69 75 77 78 79
We describe the judgment criteria, the labeling process, and the evaluation metrics in this section.

\subsection{Assessment Guidelines.}

The evaluation of the returned passages is similar to relevance assessment in other TREC settings. However, the conversational setting introduces several unique challenges.
\begin{enumerate}
    \item \textit{Contextualized}: The meaning of a turn and the relevance of an answer passage may depend on preceding turns in the same conversation. For example, “What is throat cancer?” followed by “Is it treatable?”  Each question must be interpreted in the context established by the preceding turn.
    \item \textit{Coreference and omission}: As with most human conversations, many CAsT turns have some form of ellipsis, for example, pronouns and implied context that omits words that can be understood from the preceding context.  To aid assessment, the track organizers provide resolved versions of each turn.  For example, the resolved version of “Is it treatable?” is rewritten to “Is throat cancer treatable?”.
    \item  \textit{Brevity and Completeness}: Conversational assistants interact with users via spoken or chat interfaces that are designed for brief responses.  The answer passages in the CAsT corpus tend to be short.  A good system will select passages that provide a complete answer in a concise response.
\end{enumerate}

The relevance standard for a [turn, passage] pair is intended to represent how a person would feel if she asked the question to her favorite conversational assistant (Siri, Cortana, Alexa, Google Assistant, etc.) and it responded with the passage.
A five-point relevance scale from the Google Needs Met rating scale\footnote{\url{https://static.googleusercontent.com/media/guidelines.raterhub.com/en//searchqualityevaluatorguidelines.pdf}} was adapted for the CAsT task with the following definitions.
\begin{enumerate}
    \item \textit{Fully meets (4)}. The passage is a perfect answer for the turn. It includes all of the information needed to fully answer the turn in the conversation context. It focuses only on the subject and contains little extra information.
    \item \textit{Highly meets (3)}. The passage answers the question and is focused on the turn. It would be a satisfactory answer if Google Assistant or Alexa returned this passage in response to the query. It may contain limited extraneous information. 
    \item \textit{Moderately meets (2)}. The passage answers the turn, but is focused on other information that is unrelated to the question. The passage may contain the answer, but users will need extra effort to pick the correct portion. The passage may be relevant, but it may only partially answer the turn, missing a small aspect of the context. 
    \item \textit{Slightly meets (1)}. The passage includes some information about the turn, but does not directly answer it. Users will find some useful information in the passage that may lead to the correct answer, perhaps after additional rounds of conversation (better than nothing).
    \item \textit{Fails to meet (0)}. The passage is not relevant to the question. The passage is unrelated to the target query.  
\end{enumerate}

\subsection{Assessment Process}

The labeling approach uses the standard TREC style pooling and relevance assessments.

\textbf{Pooling.} We created the assessment pool using the two runs marked as highest priority from each participant. The pool depth was judged to depth 10. We also ensured that we included two manual runs from the organizers, but this only added a small number of results to the pool (about 30). The total pool size for all turns for the 20 target topics is 33,614 unique paragraphs, several topics were truncated early in the final assessment. 

%The average pool size for each topic is listed in Table~\ref{tab:poolstat}.

\textbf{Relevance Assessments.} The relevance assessments were done by NIST assessors during a three-week period. Six assessors each worked approximately 50 hours. The average labeling speed was about 100 minutes per turn, or about 35 second per paragraph. When labeling, the assessors were provided with both the raw utterance and also our manually re-written (``resolved'') utterance. The latter was written to contain full information to define the passage relevance without depending on previous rounds. The assessors were presented with one topic at time in order of the turns. Thus the conversational context was preserved in the labeling process.

A total 20 conversational topics are labeled (almost completely), with each of them labeled to the eighth round, on average. There are total of 173 turns judged. The statistics for the distribution of the relevance labels is provided in Table~\ref{tab:judgmentsstat}. There are on average 170 unique paragraphs per turn in the pool. Each has on average 47 paragraphs with non-zero relevance score. Turn 75\_7 had no relevant paragraphs in the assessment pool. And further turn 31\_3 had all assessed passages at least slightly relevant.

The track attracted twenty one participants with a diverse combination of techniques. This made the labeling budget a significant challenge for the first year. It was also unclear what  the optimal labeling depth and number of turns per topic should be. The evaluation results from  year one provide some observations and will help guide the design of the evaluation methodologies for year two.

 %Candidates responses will be generated by pooling the participant systems, with the possibility of including simple baselines from the organizers. The labels are expected to be labeled by NIST assessors on a standard five graded relevance scale (fails to meet, slightly meets, moderately meets, highly meets, fully meets).

\textbf{Evaluation Metrics.} 
There are two dimensions in the ranking evaluation, the ranking depth and the turn depth. 
The ranking depth is the same as for adhoc search, but we focus on the earlier positions (1, 3, 5) for the conversational scenario. The turn depth evaluates the system performance at the n-th conversational turn. 
Performing well on deeper rounds indicates better ability to understand contexts. 

We use the mean NDCG@3 as the main evaluation metric, with all conversation rounds averaged using uniform weights. We also measure the turned-depth measure based NDCG@3\&N, with the per query NDCG@3 scores averaged at depth (N).
% to evaluate the systems' ability to understand conversation contexts.
Finally we calculate the MAP and Mean Reciprocal Rank to evaluate the systems.

\begin{table*}[t]
    \centering
    \caption{Participants and their runs.}
    \label{tab:participants}
    \small
    \begin{tabular}{llcl|llcl} \hline
\textbf{Group}  & \textbf{Run ID}   & \textbf{Pooled} & \textbf{Run Type} &
\textbf{Group}  & \textbf{Run ID}   & \textbf{Pooled} & \textbf{Run Type} \\ \hline
ADAPT-DCU  	& combination	    &   & manual &          UAmsterdam 	& ilps-bert-feat1	& Y & automatic \\
ADAPT-DCU 	&  datasetreorder   &   & manual	&   UAmsterdam 	&  ilps-bert-feat2	&   & automatic \\
ADAPT-DCU 	& rerankingorder    & Y & manual &	    UAmsterdam 	& ilps-bert-featq	& Y & automatic \\
ADAPT-DCU 	&  topicturnsort    & Y & manual &          UMass 	& UMASS\_DMN\_V1	& Y & automatic \\ 
ATeam 		& humanbert	    & Y & manual &	    UMass 	&  UMASS\_DMN\_V2	& Y & automatic \\ 
ATeam 		&  pg2bert	    &   & automatic &	    USI		& bertrr\_rel\_1st	&   & automatic \\ 
ATeam 		& pgbert	    & Y & automatic &	    USI 	&  bertrr\_rel\_q	& Y & automatic \\ 
CFDA\_CLIP 	& CFDA\_CLIP\_RUN1  &   & automatic &	    USI		& galago\_rel\_1st	&   & automatic \\ 
CFDA\_CLIP 	&  CFDA\_CLIP\_RUN6 & Y & manual &	    USI		&  galago\_rel\_q	& Y & automatic \\ 
CFDA\_CLIP 	& CFDA\_CLIP\_RUN7  &   & manual &	    UvA.ILPS 	& ilps-lm-rm3-dt	& Y & automatic \\ 
CFDA\_CLIP 	&  CFDA\_CLIP\_RUN8 & Y & manual &	    VES		& VESBERT		& Y & manual \\    
CMU		& coref\_cshift	    &   & automatic &	    VES		&  VESBERT1000		& Y & manual \\	       
CMU		&  coref\_shift\_qe & Y & automatic &	    WaterlooClarke & clacBase		& Y & automatic \\	       
CMU		& ensemble	    & Y & automatic &	    WaterlooClarke &  clacBaseRerank	& Y & automatic \\ 
CMU     	&  manual\_indri    & Y & manual &	    WaterlooClarke & clacMagic		&   & automatic \\	       
ECNU-ICA 	& ECNUICA\_BERT	    &   & automatic &	    WaterlooClarke &  clacMagicRerank	&   & automatic \\ 
ECNU-ICA 	&  ECNUICA\_MIX	    & Y & automatic &	    h2oloo 	& h2oloo\_RUN2		&   & automatic \\	       
ECNU-ICA 	& ECNUICA\_ORI	    & Y & automatic &	    h2oloo 	&  h2oloo\_RUN3		& Y & automatic \\	       
ICTNET 		& ict\_wrfml	    & Y & automatic &	    h2oloo 	& h2oloo\_RUN4		& Y & automatic \\	       
RALI 		& MPgate	    & Y & automatic &	    h2oloo 	&  h2oloo\_RUN5		&   & automatic \\	       
RALI 		&  MPmlp	    & Y & automatic &	    mpi-inf-d5 	& mpi-d5\_cqw		&   & automatic \\	       
RALI 		& SMNgate	    &   & automatic &	    mpi-inf-d5 	&  mpi-d5\_igraph	& Y & automatic \\ 
RALI 		&  SMNmlp	    &   & automatic &	    mpi-inf-d5 	& mpi-d5\_intu		& Y & automatic \\	       
RUCIR 		& RUCIR-run1	    & Y & automatic &	    mpi-inf-d5 	&  mpi-d5\_union	&   & automatic \\ 
RUCIR 		&  RUCIR-run2	    & Y & automatic &	    mpii  	& mpi\_base		& Y & automatic \\	       
RUCIR 		& RUCIR-run3	    &   & automatic &	    mpii 	&  mpi\_bert		& Y & automatic \\	       
RUCIR 		&  RUCIR-run4	    &   & automatic &	    udel\_fang 	& UDInfoC\_BL		&   & automatic \\	       
RUIR 		& BM25\_BERT\_FC    & Y & automatic &	    udel\_fang 	&  UDInfoC\_TS		& Y & automatic \\	       
RUIR 		&  BM25\_BERT\_RANKF& Y & automatic &	    udel\_fang 	& UDInfoC\_TS\_2	& Y & automatic \\ 
TREMA-UNH 	& UNH-trema-ecn	    & Y & automatic &	    uogTr 	& ug\_1stprev3\_sdm	&   & automatic \\ 				       
TREMA-UNH 	&  UNH-trema-ent    & Y & automatic &	    uogTr 	& ug\_cedr\_rerank	& Y & automatic \\ 
TREMA-UNH 	& unh-trema-relco   &   & automatic &	    uogTr 	&  ug\_cont\_lin	& Y & automatic \\ 
		& 		    &   &           &	    uogTr	& ug\_cur\_sdm 		& Y & manual \\	 \hline      
    \end{tabular}
    
\end{table*}

\section{Participants}

CAsT received 65 run submissions from 21 teams shown in Table ~\ref{tab:participants}. This includes 2 teams and 8 submissions from the organizing institutions. When submitting, we asked the participants to provide metadata describing certain properties of their runs. 

%The statistics of the submitted runs and the pooled runs (top two runs per team + two organizer manual runs) are listed in Table~\ref{tab:runmeta}.

\subsection{Submitted team descriptions}
Below are brief summaries of approaches from each participant. Teams are listed in alphabetical order.
\begin{itemize}

\item \textbf{ATeam} We trained several sequence-to-sequence generation models to translate questions augmented with previous conversation turns into stand-alone questions that are afterwards used to retrieve relevant passages with Anserini and re-rank them using a BERT-based model. Our question rewriting approach follows the transfer learning paradigm in which we utilised a pre-trained GPT-2 Transformer model and fine-tuned on the question rewriting task using a newly developed conversational QA dataset.

\item \textbf{ADAPT-DCU} We focus finding relevant information using contextual information from the queries. We divide our investigation of finding relevant information for conversational search into two aspects: i) Effective query formulation using syntactic analysis, ii) Data Fusion results for re-ranking top candidates retrieved from three different data sources.

\item \textbf{CFDA\_CLIP \& h2oloo} The core methods uses BM25 retrieval, doc2query to expand the MSMARCO paragraphs, and reranking use a BERT Model trained on MS MARCO. We propose two ad-hoc and intuitive approaches: Historical Query Expansion and Historical Answer Expansion, to improve the performance of the conversational IR system with limited training data.

\item \textbf{CMU} CMU used BERT attention features for coreference resolution, and identifies context shift using KL Divergence between top retrieved documents for each turn in the conversation. Retrieval is done using Indri with (and without) query expansion.

\item \textbf{ECNU-ICA} Developed a retrieval-based conversational systems named \textsc{linber}. \textsc{linber} features include five modules: coreference resolution, keywords extraction, entity linking, retrieval using Elastic Search, and BERT re-ranking.

\item \textbf{mpi-inf-d5} We propose an unsupervised method, termed CROWN: Conversational passage ranking by Reasoning Over Word Networks. CROWN works by formulating the passage score for a query as a combination of similarity and coherence, where this score is the objective to be maximized. CROWN builds a word-proximity network (WPN) from a large corpus, where words are nodes and there is an edge between two nodes if they co-occur in the same passage in a statistically significant way. 
%At query time, the WPN is used to rank passages by restricting terms (boosted via node weights) and term pairs of relevance to those that appear in the query, and that have an edge in the network. 
Finally, passages are ranked using a weighted combination of the Indri retrieval score, a node score, and an edge score.

\item \textbf{mpii} Our approach consists of an initial stage ranker followed by a BERT-based neural document re-ranking model. BM25 with query expansion
based on external knowledge (i.e., Wikipedia and ConceptNet) serves as
the first stage ranking method, while the neural model uses BERT
embeddings and a kernel-based ranking module (K-NRM) to predict a
document-query relevance score. For training we repurpose and rewrite subtopics from the TREC Web Track's diversity task in such a way that the diversity task's existing relevance
judgments may be used.

\item \textbf{RUCIR} Methods vary widely across runs and use a variety of text matching and learning to rank frameworks. AllenNLP is used for coreference resolution as well as key-value memory networks. Ranking is performed with a KNRM model as well as a MLP. A final query is generated and ranked using Indri. 

\item \textbf{RUIR} The Radboud University IR team (RUIR) investigated the usefulness of conversation context for ranking. We first rank all passages based on the union of all the turns in the conversation, using BM25, to create a pool of candidate answers. 
Next, we rerank the pool using BERT for 1) just the final question in the conversation, or 2) taking the max score fusion results of reranking with the three last questions in the conversation. The method had been tuned on the MS MARCO passage collection using different amounts of context and varying rank and score fusion methods. 

\item \textbf{TREMA-UNH} Our methods are based on entity and passage features without any dedicated question answering and dialogue tracking component. The base run works on entity relations that co-occur in top ranked passages. Subsequent runs build on that and combine both text and entity features. 

%Our first run which is based on entity relations serves as input for the second and third runs which explores both text and entity features. For the first run, we consider a relation between two entities if they co-occur in query-relevant feedback passages and measure each co-occurring entity-pair through the rank of feedback passage. We score each entity by accumulating the entity- pairs' score and select top 100 entities from this ranking. Using these ranked entities, we re-rank the feedback passages by adding the entities score to the retrieval score of feedback passages. For the second run, we concatenate all passages mentioning an entity and derive a distribution over the other entities in this composite document. We use this to rank the co-occurring entities and then use the scores of these co-occurring entities to score the passages by accumulating the entity scores of entities mentioned in the passage. For the third run, we rank passages by the number of entities in the passage which are also retrieved for the query.

\item \textbf{udel\_fang} We proposed a method that consists of two key steps: query formatting and passage re-ranking. We first apply the coreference resolution model and add the topic title. We retrieve the top 100 passages with Indri and re-rank them in a second phase using a fine-tuned BERT model.

\item \textbf{uogTr} Glasgow used probabilistic retrieval based on the Sequential Dependence model. Experiments were performed varying previous turns as weighted context as well as combined with feedback models (RM3).  A model was also trained that used the CEDR deep learning model trained on MS MARCO for reranking passages.

\item \textbf{UMass} Our re-ranking model is based on convolutional neural networks. It takes  into account the context and benefits from bag-of-words pre-trained embeddings (i.e., word2vec). Interaction matrices 
are fed to a CNN followed by max pooling, and then a  BiGRU layer. We 
finally feed the learned representation to a multi-layer perceptron 
(MLP) network to generate the matching score. We train the model on MS 
MARCO session data in a pairwise setting.

\item \textbf{USI} To understand the dependency of conversation turns, we have annotated each turn of the conversation with related (relevant) turns in the conversation’s context. We employed a high-dimensional language and position representation-based classifier to predict the relevant utterances against current utterance(s) and use them, along with other heuristics, to reformulate the query. The passage retrieval is then performed using classical term-matching models followed by neural re-ranking.

\item \textbf{UvA.ILPS} Submitted both unsupervised and supervised approaches. The unsupervised run is based on language modeling and expansion using relevance feedback (RM3). Our supervised runs rerank the set of passages retrieved by the unsupervised run. BERT is used to encode the sequence of queries up to the current turn and the passage to produce a matching score. The final matching score is obtained by linearly combining the BERT score and the unsupervised ranker's score.  MS MARCO was used for pretraining as well as a dataset originally proposed for a different task (Question Answering in Context). 

\item \textbf{WaterlooClarke} The overall approach can be explained as three steps:  1) query construction, 2) passage retrieval and ranking, and 3) passage re-ranking. Query construction used crude methods to improve retrieval performance and maintain conversational context between turns. Passage retrieval and ranking used standard BM25 ranking with pseudo-relevance feedback. Re-ranking was treated as a classification task with class probabilities used for re-ranking.

\end{itemize}
%\subsection{Run Features}
%\jeff{cx - Insert a table of the run features from the survey.}

We observe diverse approaches being utilized. There are traditional retrieval based methods, feature based learning-to-rank, neural models, and knowledge enhanced methods. A common theme through many of them is the use of BERT-based reranking methods.

\section{Overall Results}
In this section we present results of the submitted runs.

\begin{table}
\centering
\caption{Automatic response retrieval results.}
 \label{tab:automatic-results}
  \resizebox{\columnwidth}{!}{%
\begin{tabular}{l|l|c|c|c}
\hline
\textbf{Run} & \textbf{Group} & \textbf{MAP} & \textbf{MRR} & \textbf{NDCG@3}\\
\hline
SMNgate & RALI & 0.030 & 0.072 & 0.008\\
ECNUICA\_BERT & ECNU-ICA & 0.008 & 0.106 & 0.021\\
mpi-d5\_union & mpi-inf-d5 & 0.098 & 0.274 & 0.078\\
MPmlp & RALI & 0.054 & 0.285 & 0.090\\
SMNmlp &RALI & 0.060 & 0.244 & 0.090\\
UMASS\_DMN\_V1 & UMass & 0.077 & 0.298 & 0.108\\
MPgate & RALI & 0.053 & 0.282 & 0.108\\
\textit{indri\_ql\_baseline} & - & 0.139 & 0.328 & 0.152\\
galago\_rel\_q & USI & 0.105 & 0.394 & 0.181\\
galago\_rel\_1st & USI & 0.112 & 0.426 & 0.197\\
ECNUICA\_MIX & ECNU-ICA & 0.171 & 0.522 & 0.231\\
mpi\_base & mpii & 0.173 & 0.508 & 0.234\\
ECNUICA\_ORI & ECNU-ICA & 0.190 & 0.519 & 0.242\\
RUCIR-run2 & RUCIR & 0.092 & 0.494 & 0.253\\
UDInfoC\_TS\_2 & udel\_fang & 0.061 & 0.541 & 0.253\\
coref\_cshift & CMU & 0.213 & 0.505 & 0.253\\
RUCIR-run3 & RUCIR & 0.093 & 0.502 & 0.255\\
ilps-lm-rm3-dt & UvA.ILPS & 0.229 & 0.528 & 0.267\\
coref\_shift\_qe & CMU & 0.224 & 0.509 & 0.272\\
RUCIR-run4 & RUCIR & 0.105 & 0.527 & 0.273\\
UDInfoC\_TS & udel\_fang & 0.067 & 0.567 & 0.278\\
mpi-d5\_cqw & mpi-inf-d5 & 0.185 & 0.591 & 0.286\\
mpi-d5\_igraph & mpi-inf-d5 & 0.187 & 0.597 & 0.287\\
mpi-d5\_intu & mpi-inf-d5 & 0.240 & 0.596 & 0.289\\
ensemble & CMU & 0.258 & 0.587 & 0.294\\
bertrr\_rel\_q & USI & 0.141 & 0.516 & 0.298\\
bertrr\_rel\_1st & USI &0.146 & 0.539 & 0.308\\
UDInfoC\_BL & udel\_fang & 0.075 & 0.596 & 0.316\\
mpi\_bert & mpii & 0.166 & 0.597 & 0.319\\
ug\_cont\_lin & uogTr & 0.275 & 0.584 & 0.325\\
ug\_1stprev3\_sdm & uogTr & 0.253 & 0.585 & 0.328\\
clacBaseRerank & WaterlooClarke & 0.244 & 0.629 & 0.343\\
BM25\_BERT\_RANKF & RUIR & 0.158 & 0.597 & 0.350\\
ilps-bert-feat2 & UAmsterdam & 0.256 & 0.603 & 0.352\\
BM25\_BERT\_FC & RUIR & 0.158 & 0.601 & 0.354\\
ug\_cedr\_rerank & uogTr & 0.216 & 0.643 & 0.356\\
clacBase & WaterlooClarke & 0.246 & 0.640 & 0.360\\
ilps-bert-featq & UAmsterdam & 0.262 & 0.653 & 0.365\\
ilps-bert-feat1 & UAmsterdam & 0.260 & 0.614 & 0.377\\
pg2bert & ATeam & 0.258 & 0.641 & 0.389\\
pgbert & ATeam & 0.269 & 0.665 & 0.413\\
h2oloo\_RUN2 & h2oloo & 0.273 & 0.714 & 0.434\\
CFDA\_CLIP\_RUN7 & CFDA\_CLIP & 0.267 & 0.715 & 0.436\\
\hline
\end{tabular}%
}

\end{table}

\begin{table}
\caption{Manual response retrieval results. These runs used the manually resolved queries.}
 \label{tab:manual-results}
 \resizebox{\columnwidth}{!}{%
\begin{tabular}{l|l|c|c|c}
\hline
\textbf{Run} & \textbf{Group} & \textbf{MAP} & \textbf{MRR} & \textbf{NDCG@3}\\
\hline
UMASS\_DMN\_V2 & UMass & 0.082 & 0.300 & 0.100\\
ict\_wrfml & ICTNET & 0.105 & 0.373 & 0.165\\
UNH-trema-ecn & TREMA-UNH & 0.073 & 0.505 & 0.222\\
unh-trema-relco & TREMA-UNH & 0.077 & 0.533 & 0.239\\
UNH-trema-ent & TREMA-UNH & 0.076 & 0.534 & 0.242\\
topicturnsort & ADAPT-DCU & 0.136 & 0.555 & 0.259\\
rerankingorder & ADAPT-DCU & 0.137 & 0.564 & 0.259\\
combination & ADAPT-DCU & 0.130 & 0.539 & 0.259\\
datasetreorder & ADAPT-DCU & 0.135 & 0.550 & 0.260\\
VESBERT & VES & 0.124 & 0.541 & 0.291\\
VESBERT1000 & VES & 0.204 & 0.555 & 0.304\\
\textit{manual\_indri\_ql} & - & 0.309 & 0.660 & 0.361\\
clacMagic & WaterlooClarke & 0.302 & 0.687 & 0.411\\
clacMagicRerank & WaterlooClarke & 0.301 & 0.732 & 0.411\\
RUCIR-run1 & RUCIR & 0.163 & 0.725 & 0.415\\
ug\_cur\_sdm & uogTr & 0.334 & 0.715 & 0.421\\
CFDA\_CLIP\_RUN1 & CFDA\_CLIP & 0.224 & 0.772 & 0.460\\
h2oloo\_RUN4 & h2oloo & 0.319 & 0.811 & 0.529\\
h2oloo\_RUN3 & h2oloo & 0.322 & 0.810 & 0.531\\
CFDA\_CLIP\_RUN8 & CFDA\_CLIP & 0.361 & 0.854 & 0.560\\
h2oloo\_RUN5 & h2oloo & 0.352 & 0.864 & 0.561\\
CFDA\_CLIP\_RUN6 & CFDA\_CLIP & 0.392 & 0.861 & 0.572\\
humanbert & ATeam & 0.405 & 0.879 & 0.589\\
\hline
\end{tabular}%
}

\end{table}

We first present results macro-averaged at the level of every turn independently. We use three standard TREC evaluation measures, Mean-average Precision (MAP), and Normalized Discounted Cumulative Gain (NDCG), and Mean Reciprocal Rank (MRR).  In particular, we use NDCG@3 as the primary measure because we focus on graded relevance of results at the top ranks. 

For reporting results we make a distinction between \textit{automatic} and \textit{manual} runs. Automatic runs use the provided test topics. Manual runs use the test topics, but use the manually rewritten (resolved) queries where coreference and other phenomena have been replaced to create clear and unambiguous utterances.  

\textbf{Automatic run results.} The results for the 41 automatic runs are provided in Table~\ref{tab:automatic-results}. The results show systems that vary widely in effectiveness. The median NDCG@3 score of the automatic runs is 0.286. The best performing run not utilizing BERT for reranking is clacBase with a NDCG@3 value of 0.360. We observe that nine of the top ten best performing runs use a form of BERT for ranking results. The top two performing teams perform contextual query rewriting and expansion.

\textbf{Manual run results.} The results for the 24 manual runs are provided in Table~\ref{tab:manual-results}. The median manual run has an NDCG@3 value of 0.361. We observe that the manual\_indri query-likelihood run provided by the organizers is the median run. 
%The other organizer run of ug\_cur\_sdm is a sequential dependence model that outperforms all non-BERT (and some BERT) reranking models by NDCG@3 (although it is lower on MRR). 
This indicates that models trained on the limited training data may not have generalized well. The best performing run is humanbert with an NDCG@3 value of 0.589. The humanbert run uses BERT-large as a reranking method on top of Anserini results. Similar to the automatic runs, the best performing runs all leverage BERT as a feature in reranking. We also observe the gap between the best manual and automatic runs is large, a 26\% relative difference in median and 35\% relative difference in the best runs. 

%NDCG@3 by turn depth
%0.299
%0.249
%0.309
%0.300
%0.257
%0.263
%0.295
%0.229
%0.254
%0.270
%0.273

\textbf{Influence of WaPo posting filtering.} Post-filtering the Washington Post paragraphs in the run results may benefit those runs do not include WaPo results. There are eight such runs and four are top performing ones: BM25\_BERT\_RANKF, BM25\_BERT\_FC, pgbert, and humanbert. Overall about 10\% of pool candidates are from WaPo.

\textbf{Pool Incompleteness.} Due to assessment resource constraints, only the top two prioritized runs (out of total maximum four) from each group are pooled. The influence of this incomplete pooling appears to be small for submitted systems. There is about a 10\% absolute NDCG score difference between pooled and unpooled runs; however, it is not clear how much of this gap originates from the preference of the teams---they likely picked their best runs to be in the pool. There are on average 0.6 passages unjudged from top runs (in the top 10). The average number of unjudged documents in the top 10 results is 1.67 per turn for all unpooled runs. Approximately half of the unjudged documents are from three groups with the lowest effectiveness overall. Further, the top performing run is not included in the pool. Although more work could be done, we are optimistic about the reusability of the produced benchmark.

\subsection{Results by turn depth}

\begin{figure}
\centering
\includegraphics[width=0.5\textwidth]{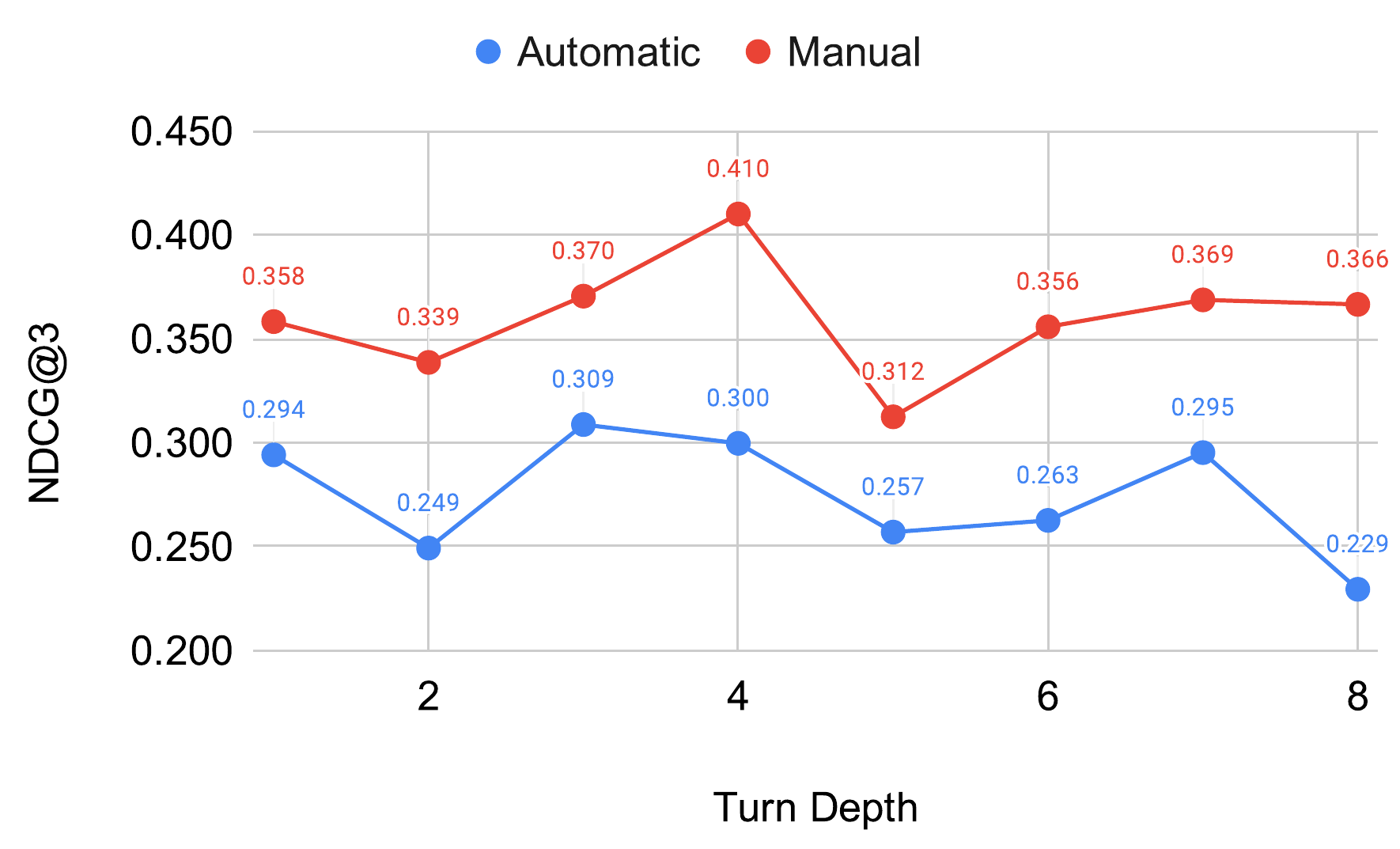}
\caption{NDCG@3 at varying conversation turn depth.}
\label{fig:turn-depth}
\end{figure}
We plot the average NDCG@3 at each turn depth. The results are shown in Figure~\ref{fig:turn-depth}. Turns beyond eight are truncated due to small sample size. We measure statistical significance with paired t-test at greater than 95\% confidence interval (with Bonferroni correction where needed).

For automatic runs the table shows the best depths are early at 1 and 3. For an unknown reason there is a significant dip for automatic runs at the second turn. There is a downward trend from an average of approximately 0.3 at the first turn to an average of 0.23 by turn eight. This represents a statistically significant decrease of 23\% from the start turn to the end. However, the variation in the decline varies by depth. For example, depth 7 is statistically equivalent to the effectiveness of the first turn. The reason for this behavior bears further investigation. 

The results for manual runs show a different pattern. Effectiveness dips very slightly (insignificantly) at turn 2, but increases steadily until turn 4. It drops at turn 5, possibly due to shifting subtopics. Unlike for the automatic runs the result at the end of the conversation is statistically equivalent to the start of the conversation. The manually resolved queries do not face the challenge of conversational query rewriting. When compared with the automatic runs the results show an increasing gap in system effectiveness over time (except at 5 where both perform poorly). Comparing the start and the end there is a more than 100\% relative increase in the effectiveness gap between manual and automatic runs.

\section{Results Analyses}
We encouraged the groups to submit metadata along with their runs. We designed a questionnaire with Yes/No questions in three categories: Query Understanding, Training/Retrieval Models, and the Utilization of Context information. Each question asks the teams to provide whether their runs used a specific type of resource or technique. This section discusses the impact of the varying approaches and resource usage on system effectiveness. 

\subsection{Query Understanding}
\begin{figure}
\centering
\includegraphics[width=0.45\textwidth]{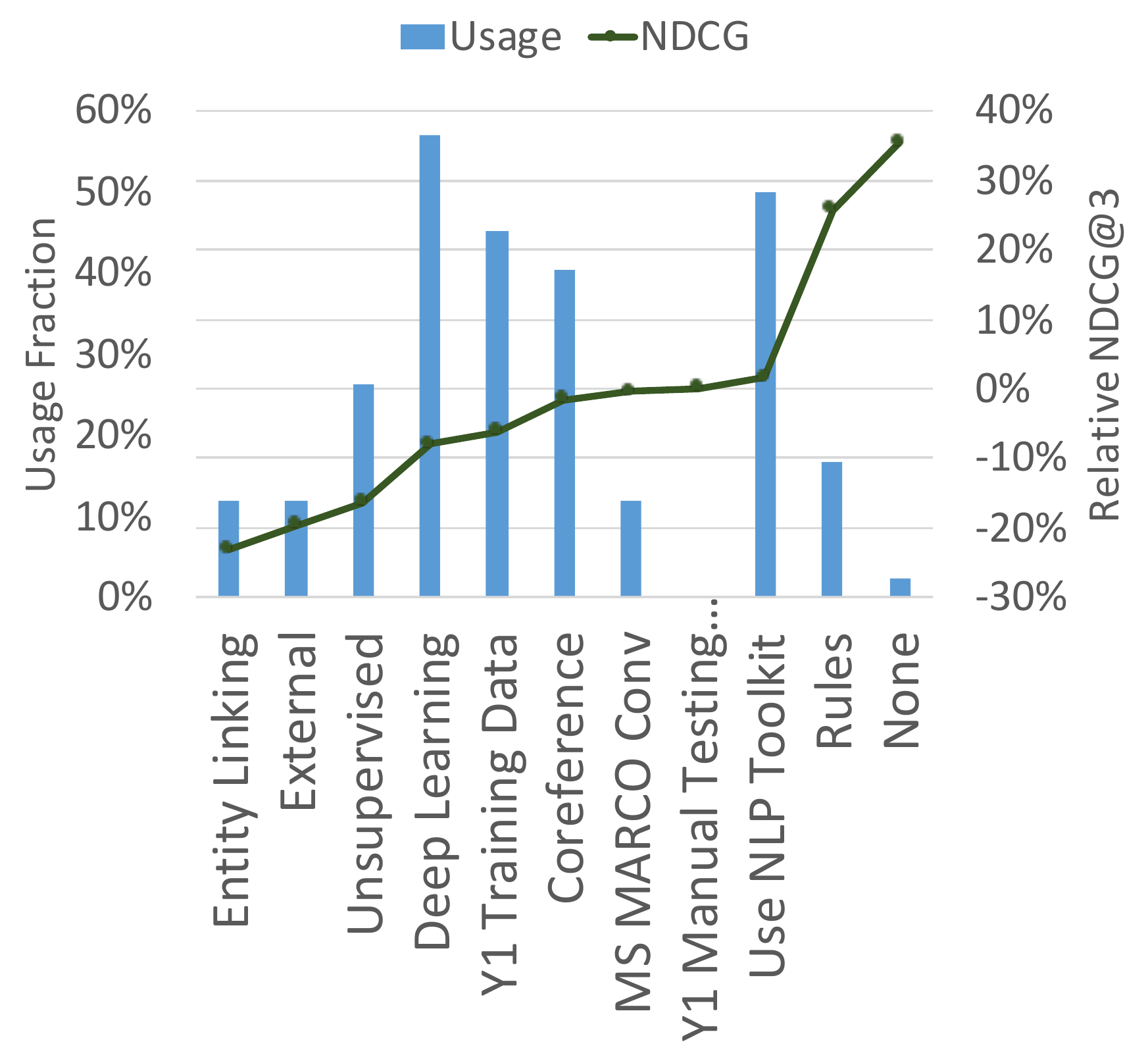}
\caption{Query understanding influences on system effectiveness for automatic runs.}
\label{fig:meta-influences-query}
\end{figure}

\begin{table*}[th]
    \centering
    \caption{Query understanding examples from four reference types.     \label{tab:nlueg}}

    \begin{tabular}{l|l|l}
    \hline \hline
    \textbf{Type}     &  \textbf{Utterance} & \textbf{Understanding} \\ \hline
    Pronominal & How to \textit{they} celebrate Three Kings Day? & they $\rightarrow$ Spanish People \\ 
    Zero & What cakes are traditional? & Null$\rightarrow$Spanish, Three Kings Day \\
    Groups & Which \textit{team} came  first? & team$\rightarrow$Avengers, Justice League \\
    Abbreviations & What are the main types of VMs? & VMs$\rightarrow$Virtual Machines \\ \hline \hline
    \end{tabular}
\end{table*}
\begin{table}[t]
    \centering
      \caption{Manual counts of four reference types in CAsT Y1 (2019) conversational topics.  \label{tab:nlustat}}
    \begin{tabular}{l|r|r}
    \hline \hline
       \textbf{Type}  & \textbf{Train} & \textbf{Test}  \\ \hline
    Pronominal     &  102 & 128 \\
    Zero & 82 & 111 \\
    Groups & 6 & 4 \\ 
    Abbreviations & 29 & 15 \\ \hline \hline
    \end{tabular}

\end{table}

The query understanding questions show the techniques  used in understanding the conversational queries, including but not limited to query rewriting, term re-weighting, query expansion, coreference resolution, and others.

In total the following eleven binary questions are asked.
\begin{enumerate}
    \item \textit{Entity Linking.} Whether the method uses entity linking techniques on the query.
    \item \textit{External.} Whether the method uses external data.
    \item \textit{Unsupervised.} Whether the method uses unsupervised query understanding technique.
    \item \textit{Deep} \textit{Learning.} Whether the method uses deep learning.
    \item \textit{Y1} \textit{Training.} Whether the method uses CAsT Y1 training dataset.
    \item \textit{Coreference.} Whether the method uses coreference resolution.
    \item \textit{MS} \textit{MARCO} \textit{Conv.} Whether the method uses data from the MS MARCO Conversational Session dataset.
    \item \textit{Y1 Manual Testing Query Annotation.} Whether the method uses CAsT Y1 annotated (resolved) query dataset.
    \item \textit{NLP} \textit{Toolkit.} Whether the method uses a standard NLP toolkit.
    \item \textit{Rules.} Whether the method uses heuristic rules.
    \item \textit{None.} No query understanding method is used.
\end{enumerate}

For our analysis we focus on automatic runs because manual runs had key understanding issues removed. Figure \ref{fig:meta-influences-query} plots he fraction of the automatic runs using each feature (answering Yes to the question) as well as the relative NDCG@3 performances of those using the feature over those that do not.

The most popular technique in query understanding is deep learning, with 57\% of runs using it, but the influence is slightly negative, performing 8\% worse on average than those runs not using it. NLP toolkits are used by half of runs but there is no relative gain by using it. 
There are only two runs labeled as ``None'' in the query understanding category, so its relative gain might not be reliable due to small sample size (one run was the best performing CFDA\_CLIP\_RUN7 run.  

Overall the results reveal the challenge of query understanding in CIS. 
Many widely utilized techniques from adhoc retrieval led to negative \textit{average} contribution to conversational search accuracy. The only significant gains are from manually designed rules that used query term reweighting and conversational stopword removal. The most effective method was a form of query expansion leveraging results from previous turns.

To characterize the nature of the query understanding challenges the organizers manually analyzed the topics. We highlight four primary types of language coreference phenomena observed. Examples of the four phenomena and their statistics in the CAsT Y1 train and test conversations are listed in Table~\ref{tab:nlueg} and Table~\ref{tab:nlustat}. How to handle these types of conversational contextualization effectively is one of the main challenges for CIS identified in CAsT Y1. In particular, the use of zero anaphora (implied mention) and group coreference differs from typical coreference in other text genres.

\subsection{Retrieval and Ranking}
\begin{figure}
\centering
\includegraphics[width=0.5\textwidth]{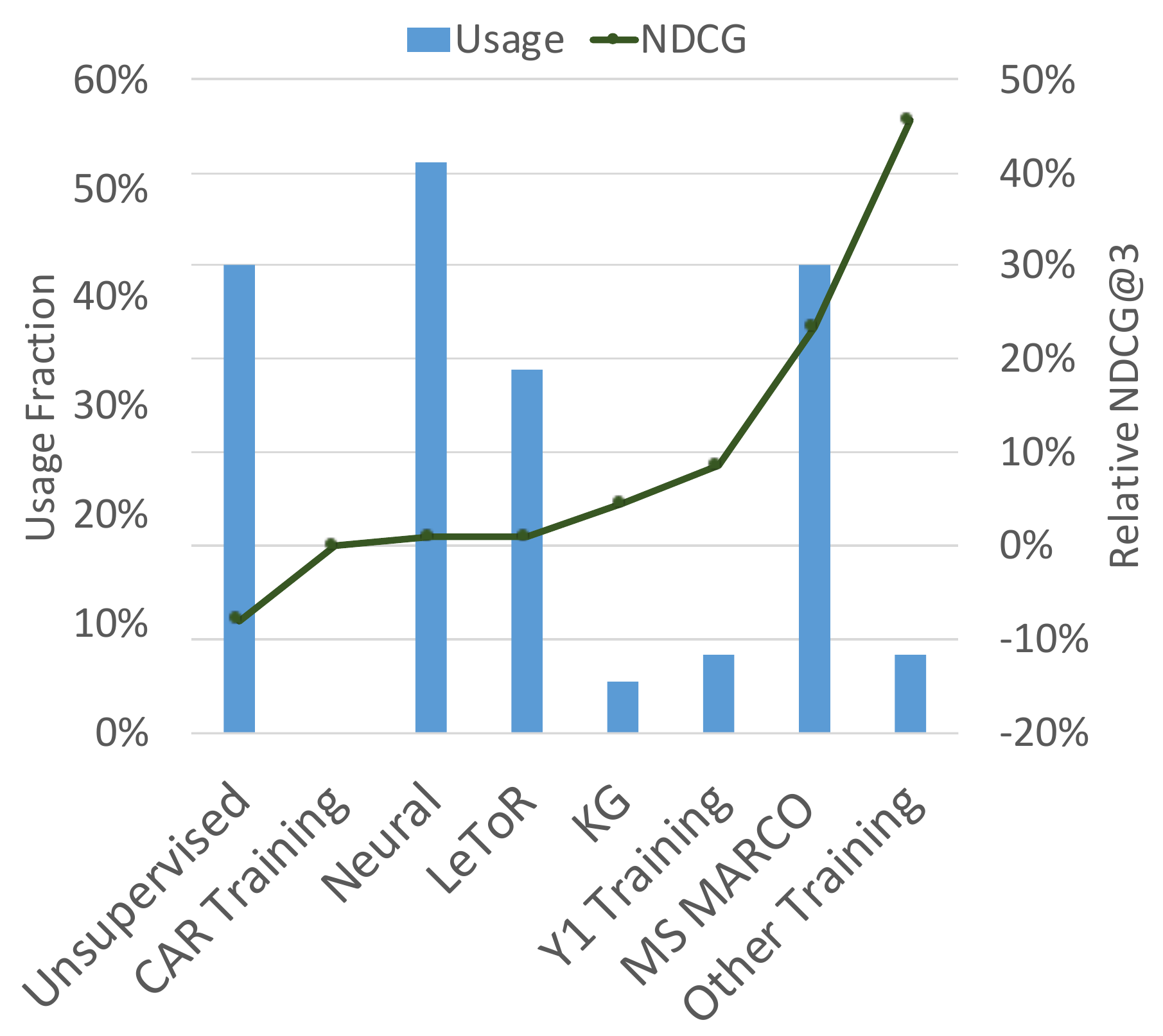}
\caption{Training and retrieval model influences on system effectiveness for automatic runs.}
\label{fig:influences-retrieval}
\end{figure}

Eight questions were asked on the data and techniques used in the retrieval and ranking model components.
\begin{enumerate}
    \item \textit{Unsupervised.} Whether any training data has been used.
    \item \textit{CAR Training.} Whether it is trained using TREC CAR dataset.
    \item \textit{Neural.} Whether deep learning is used.
    \item \textit{LeToR.} Whether learning-to-rank is used.
    \item \textit{KG.} Whether the method uses a knowledge graph.
    \item \textit{Y1 Training.} Whether the method is trained with CAsT Y1 training data.
    \item \textit{MS MARCO Training.} Whether the method is trained with MS MARCO dataset.
    \item \textit{Other Training.} Whether the method is trained with other datasets.
\end{enumerate}
The fraction of each technique/data being used and their influence for automatic runs is shown in Figure ~\ref{fig:influences-retrieval}.  

Utilizing any types of training data leads to improvements in effectiveness. The unsupervised runs, consisting of 43\% of all runs, performed 8\% worse on average. The MS MARCO training data is the most frequently used supervision source. It was also the recommended source for single turn relevance training data. Using it leads to a greater than 20\% improvement on average. This is not surprising given the large fraction of MARCO passages in our corpus and the similarity between the single turn passage ranking task. The other most effective method is to leverage `other' additional training data, although the sample size is small; it was only used by four runs from two teams. In particular, this reflects the three strongly performing ATeam runs that used Google Natural Questions as a training source. 

The most used ranking method is neural ranking methods, half of the CAsT Y1 runs leveraged deep learning techniques. The effectiveness of using them, however, is mixed. On average, using deep learning does not necessary lead to better ranking effectiveness. There is no difference over teams not using them on average. However, at the same time, nine out of top ten best performing runs used neural approaches. This shows the potential of these methods as well as the challenge in using them with consistent effectiveness. 

Figure ~\ref{fig:influences-retrieval} only reports data for automatic runs, but the results across all runs are similar. The biggest difference between automatic and manual runs is the effectiveness of neural methods. For manual runs they show a 19\% relative improvement versus 1\% for automatic runs. The current neural methods appear to be more effective on reranking for the manually resolved queries, which bears further investigation. It could be due to improved result recall, the resolved queries themselves, or combination of both. 

\subsection{Conversational Context}
\begin{figure}
\centering
\includegraphics[width=0.5\textwidth]{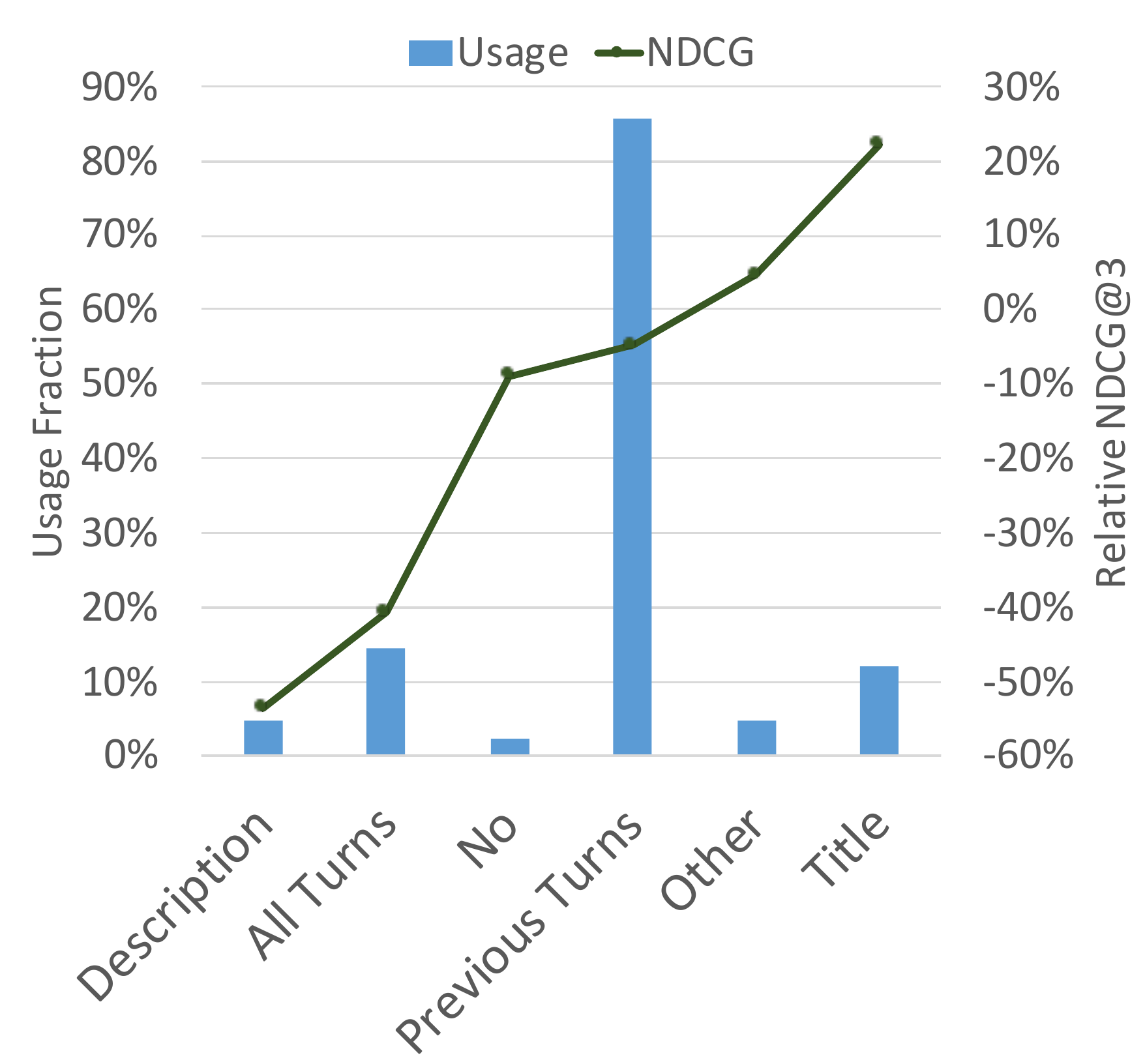}
\caption{Retrieval context influences on system effectiveness for automatic runs.}
\label{fig:influences-context}
\end{figure}
Another key challenge in conversational search is to accurately infer information needs through the \textit{conversation history} in the multi-round interactions. The inferred user's information needs can be used as part of the query understanding in query expansion/rewriting, or in the ranking models. 

There are a variety of contextual information a CIS system could leverage. In CAsT Y1, we asked about five types of context.
\begin{enumerate}
    \item \textit{Description.} The long description of the topic.
    \item \textit{All Turns.} The utterances in the conversation topic, both before and after the current turn. 
    \item \textit{Previous Turns.} The utterances before the current turn.
    \item \textit{Other.} External information outside those provided by CAsT.
    \item \textit{Title.} The short keyword title of the topic.
    \item \textit{No.} No context information is used, only the current turn.
\end{enumerate}
Figure~\ref{fig:influences-context} shows the use of the varying context information and their influence on automatic runs.

The majority of runs (86\%) utilized previous turns in the multi-round conversation; these are crucial to resolve the contextual dependence of the current turn using previous information. The title of the conversation topic is the most effective context; it is manually written by the organizers and its keyword-like style can be effectively handled by adhoc retrieval systems. 

A few runs used the description or all turns of the conversation. These two resources are challenging to utilize. For all turns, we also think this metadata is be noisy and some teams may have interpreted all turns as previous turns. For the description, a potential reason could be that the long description is difficult to use effectively. 

For the manual runs, a much smaller fraction used previous turn context,  only 20\%. This is because ambiguous context was manually rewritten. 

The descriptions, latter turns, and conversation titles are not actual interactions between the user and the CIS system. These are unlikely to be available in actual conversational search systems and future iterations may not allow their use by automatic systems.

\section{Conclusion}
In the first year of TREC CAsT we learned a lot about the structure of the problem of conversational search. 
\begin{itemize}
\item \textbf{Conversational Language Understanding.} Existing off-the-shelf coreference models struggled with TREC CAsT topics more than expected. The results using the manually resolved queries demonstrates a gap of approximately 35\% over the best automatic system.

\item \textbf{Conversational Context} The results on the manual runs show that clean context has potential to maintain or even improve effectiveness over the course of the conversation as an information need unfolds. In contrast, automatic runs show a decline in effectiveness as turn depth increases.  

\item \textbf{Ranking.} BERT-based neural models are the current leading method for response ranking across both manual and automatic methods. However, its application has mixed results. Many BERT-based runs are outperformed by simpler traditional ranking approaches. The neural reranking approaches show a larger relative gain on cleaner manually resolved queries, indicating that effective query formulation is an important factor.

%\item \textbf{Conversational structure.} Developing exploratory information topics is challenging. Developing challenging topics focusing on non-factoid questions at the appropriate length, etc... remains an open challenge.  Further the lack of result context made true conversational aspects difficult to model, e.g. system failure, etc... 

\end{itemize}
After the success of the first year we look forward to year two.

\section*{Acknowledgments}

We thank Vaibhav Kumar for his work in developing topics, training relevance assessments, and the Indri search engine provided to CAsT participants. We thank the Deep Learning track organizers for helping align the two tracks. 
We also are deeply thankful for Ellen Voorhees’ experience, patience, and persistence in running the assessment process. Finally we thank all our participants.

\bibliographystyle{splncs04}
\bibliography{bibliography}

\end{document}